\documentclass[opinion]{sigirforum}
\usepackage{array}
\usepackage{quoting}
\quotingsetup{leftmargin=20pt, rightmargin=20pt, vskip=5pt}
\usepackage[T1]{fontenc}

\newcommand{\ie}{\emph{i.e.}}
\newcommand{\eg}{\emph{e.g.}}
\newcommand{\Eg}{\emph{E.g.}}

\newcommand{\vs}{\emph{vs. }}

\def\pubissue{Vol. 60 No. 1 -- June 2026}

\begin{document}
\title{Towards Critical IR Theories and Practices}

\authors{
    \author[bhaskar.mitra@acm.org]{Bhaskar Mitra}{Independent Researcher}{Tiohtià:ke / Mooniyang / Montréal, Canada}
}

\maketitle

\abstract{
Belkin and Robertson urged us, half a century ago, to develop a theoretical foundation for understanding what constitutes \textit{societal good} that can inform information retrieval (IR) research and serve as a basis for determining when we should limit our scientific inquiry in the face of demands that are contradictory to societal good.
In this article, I argue that to achieve this, IR should embrace critical theories and practices in our work, and shift away from the dominant liberal frame through which much of the IR community today view societal
concerns in context of our research.
Unlike the liberal frame, the critical frame explicitly adopts nondomination as its stated goal which can clarify our conceptualization of societal good within the field, provide necessary theoretical underpinning that Belkin and Robertson urged the community to develop, and serve as a basis for critical appraisals of our progress in enacting desired societal change.
}

\section{Introduction}
\label{sec:intro}
Half a century ago, \citet{belkin1976some} called on the information science (IS) community to recognize the field's salient responsibilities to society.
They urged the community to develop a theoretical foundation for understanding what constitutes societal good that can inform information access research and technology development.
In particular, they wrote:
\begin{quoting}
  ``\emph{Thus, information science must become theoretically self-conscious, and self-consciously based upon a social ideology. Inevitably such a position will lead to conflicts between ethical and politically expedient imperatives. Not only must we be willing to limit our scientific inquiry according to public interest, we must have a concept of the public interest and be able to argue our position when faced with contradictory demands. And in order to do this successfully, we must have developed our theoretical position and our limitations before being faced with such conflicting demands (otherwise, will we realize the conflict before it is too late?).}''
\end{quoting}
Unfortunately, in spite of their prescient call-to-action, five decades later the field of information retrieval (IR), which could be situated at the intersection of computing and IS, finds itself in exactly the troubling situation that Belkin and Robertson warned us about.
IR research and practices are often motivated by grand ambitions like ``organiz[ing] the world's information and mak[ing] it universally accessible and useful''\footnote{Source: Google's mission statement (\url{https://about.google/company-info/})} or supporting human cognition with information artifacts~\citep{zobel2018we}.
But the reality of dominant search and recommendation platforms is that they codify oppression~\citep{noble2018algorithms} and have materialized massive infrastructures for (i) large-scale user surveillance~\citep{zuboff2023age}, (ii) commodification and harvesting of user attention~\citep{wu2017attention}, and increasingly (iii) manipulating user's political and commercial preferences via generation of persuasive language and visualizations using generative AI~\citep{mitra2024sociotechnical}.
IR platforms have also increasingly become targets of ideological capture by authoritarian state and private actors because they represent sites of struggle between oppression and justice~\citep{mitra2026information}.
These concerns are particularly serious and urgent today in light of the rising levels of democratic erosion worldwide and the increasing concentration of economic and political power in the hands of Big Tech.

Simultaneously, there has been a significant uptick in IR scholarship around fairness and explainability since 2018, following the third Strategic Workshop in Information Retrieval in Lorne (SWIRL)~\citep{culpepper2018research} where IR researchers in attendance recognized fairness, accountability, confidentiality, and transparency in IR (``FACT IR'') as consequential research directions for the field.
However, how much of this research has in practice meaningfully translated into societally positive outcomes is unclear~\citep{mitra2025search}.
The current frames through which the IR community approaches societal concerns have also failed to thwart the community from engaging in research that support user surveillance and manipulation, and serious concerns about how proprietary platforms have concentrated power in the hands of few individuals and institutions making our information ecosystem vulnerable to authoritarian capture have also hitherto received insufficient attention.

The under-theorization of ``societal good'' in IR is a key contributing factor here.
After all, how can we appraise our progress in affecting positive societal change---or limit our scientific inquiry when faced with demands contradictory to societal good---if we cannot articulate what societal good we are aiming for?
This has led to several recent calls for the IR community to explicate its normative positions~\citep{vrijenhoek2023report}, values~\citep{trippas2025report}, and sociotechnical imaginaries~\citep{mitra2025search}.
Consequently, the IR-for-Good track at the 2026 edition of the European Conference on Information Retrieval (ECIR) adopted as its theme the question of ``\textit{what is IR-for-Good?}''~\citep{mitra2025ir} which again attempts to get at this exact underlying tension in the field.

Against this backdrop, this article invites the IR community to engage in thoughtful deliberation over how to develop a shared normative position on societal good.
I believe this would require IR to embrace \textit{critical} theories and practices in our work, and shift away from the dominant \textit{liberal} frame through which much of the IR community today view societal concerns in context of our ongoing research.
The liberal frame concerns itself with designing a \textit{good system} without explicating what \textit{societal good} it wants to realize in the process.
In contrast, the critical frame adopts realizing a \textit{nondominative society} as its stated goal which can clarify our conceptualization of societal good within the field, provide necessary theoretical underpinning that Belkin and Robertson urged the community to develop, and serve as a basis for critical appraisals of our progress in enacting desired societal change.
Furthermore, the critical frame grounds itself in structural understanding of intersecting systems of oppression, and extends our purview of concerns beyond how the system behaves to also include the sociopolitical context in which the system is embedded and who gets to exert power and control over the system.
The critical approach is also not constrained to exploring algorithmic mitigation exclusively and includes in its arsenal the right of refusal~\citep{barocas2020not} and to build radical alternatives.
I implore the IR community to evolve our current research practices to rigorously incorporate critical theories and perspectives, and hold ourselves accountable for our impact on society.

\section{The two frames of IR \& Society}
\label{sec:frames}
Within the relatively short period between 2018-2026, we observe that IR has employed two distinct frames---the \textit{liberal} and the \textit{critical}---to motivate societal concerns in IR research.
A frame, in this context, may be loosely defined as comprising of (i) a set of desired goals, (ii) a collection of methods and practices, and (iii) a body of underlying values, politics, and epistemologies that distinguish it from other frames.
Table~\ref{tbl:frames} summarizes some of the key distinctions between the liberal and the critical frames along these dimensions.

\begin{table}[t]
    \renewcommand{\arraystretch}{1.5}
    \begin{center}
        \small
        \captionsetup{width=0.9\linewidth}
        \caption{Contrasting the liberal and critical frames with respect to their goals, their methods and practices, and their underlying values, politics, and epistemologies.}
        \label{tbl:frames}
        \begin{tabular}{>{\raggedright\arraybackslash}p{0.16\linewidth}p{0.32\linewidth}p{0.32\linewidth}}\hline
            \textbf{} & \textbf{Liberal frame} & \textbf{Critical frame}\\\hline
            \textbf{Goals} & Desired system attributes, \eg, fairness, confidentiality, transparency, and sustainability. & Desired societal outcomes, \eg, social justice, emancipation, and democracy. \\
            \textbf{Methods and practices} & Algorithmic harm mitigation, \eg, fair ranking, fact-checking, privacy mechanisms, transparency, and explainability. & Sociotechnical mitigation, repair, refusal, building counter-technologies; foregrounds praxis and movement building. \\
            \textbf{Values, politics, and epistemologies} & Sometimes (mis-)represented as apolitical; politically espouses individual rights, equality, freedom of speech, and market economies; philosophically positivist and tied to technological and instrumental rationality. & Explicitly political and politically decolonial, anti-racist, feminist, queer, pro-labor, anti-capitalist, anti-casteist, anti-ableist, and abolitionist; philosophically social constructionist and principally nondominative. \\\hline
        \end{tabular}
	\end{center}
\end{table}

\subsection{The liberal frame}
\label{sec:frames-liberal}
The liberal frame emphasizes desired attributes of the IR  system---\eg, fairness, confidentiality, transparency, and sustainability---and consequently focuses on algorithmic mechanisms to achieve these objectives.
We can see this reflected in the call for ``FACT IR'' at SWIRL 2018~\citep{culpepper2018research}, and in the technical research agenda proposed for Responsible IR by the FACTS-IR workshop~\citep{olteanu2021facts} organized the following year.
Subsequently, there has been a broad body of work on fairness~\citep{ekstrand2022fairness, zehlike2021fairness, patro2022fair, wang2023survey, li2023fairness}, explainability~\citep{zhang2020explainable, anand2022explainable}, and other related topics in IR.

The liberal frame largely distances itself from any explicit political alignments.
But the absence of political expression does not imply an absence of politics in the research conducted through this frame.
Instead, the liberal frame maybe better understood as valuing individual rights, equality, freedom of speech, and market economies.
IR has historically held positivist notions about the ``neutrality'' of information and its research practices, and operated within the frames of technological and instrumental rationality~\citep{pyati2006critical}.
The liberal frame also draws from these IR traditions and is grounded in related epistemologies.

By defining its goals in terms of what constitutes a \textit{good system}, the liberal frame subtly sidesteps the question of what constitutes a \textit{good society}.
The latter questions is saliently more political and I speculate that posing it might be a source of much discordance within the IR community whose members reflect vastly different values, cultures, politics, identities, privileges, lived experiences, and understandings of history.
From a cursory look, this may seem like a clever strategy to allow the community to start building a foundation for IR research that prioritizes societal good without waiting on the community to reach reasonable political consensus on what societal good we are aiming for.
However, after several years of working on fair ranking and related topics through this lens, I have reached the conclusion that any attempt to bypass or depoliticize the question of what constitutes societal good is ultimately self-defeating and cannot lead us towards ensuring positive societal impact.

The challenges of defining a good system without defining a good society is multifold.
Let's take fairness in IR, for example.
It is reasonable to assume that a majority of the IR community prefer that our IR systems be fair.
However, how we can move fairness from an ideal to a practical implementation in the context of historical dimensions of marginalization---such as race and gender---is still unclear in spite of the substantive body of existing research in this area.
A fundamental challenge is that racial and gender categories are multidimensional, relational, and socially-constructed.
However, as \citet{hanna2020towards} and \citet{pinney2023much} point out, typical fairness frameworks  represent these categories as single-dimensional variables that can take one of few discrete values and treat groups as interchangeable erasing the context of historical harms and the distinct nature of marginalization faced by different groups.
As \citet{hanna2020towards} summarize: ``In short, group fairness approaches try to achieve sameness across groups without regard for the difference between the groups.
Group fairness offers an incomplete version of what a race-conscious policy would be.''
Practical deployment of fairness interventions also raises concerns of further intensification of data collection on marginalized groups and creation of population-scale databases containing categorical data on algorithmic subjects that risk being instrumentalized by state actors and other powerful entities to target historically marginalized groups.

Fairness constructs in IR also typically treat information artifacts as interchangeable commodities valued through the lens of relevance, source authority, factuality, and other similar factors with little regard to their complex relationship with societal concerns.
This is evident if we consider fairness principles in IR like \textit{equity of attention}\footnote{``\textit{A ranking offers equity of attention if each subject receives attention proportional to its relevance.}''}~\citep{asia:equity-of-attention} and \textit{equal expected exposure}\footnote{``\textit{Given a fixed information need, no item should be exposed (in expectation) more or less than any other item of the same relevance}''}~\citep{diaz2020evaluating}.
However, information artifacts are in fact \textit{not} commodities but rather objects that are defined not just by their relevance to the query but by their complex relationships with societal concerns and their role in our collective sense-making of our place in this world.
For example, say, there are two articles both deemed topically ``relevant'' to the topic of the civil rights movement in the US but one article is grounded in anti-racist epistemology while the other espouses white supremacist point-of-view; surely these two articles do not deserve equal exposure.
This example fits into a broader pattern of glaring misapplication of the bias and fairness construct in IR in the context of political ideologies.
There are several examples of IR papers\footnote{I intentionally do not list specific papers as examples here because my goal is not to shame the authors, but to grow our collective sociopolitical sensitivity to these concerns.} that study whether specific IR methods give disparate representation to different ends of the political spectrum (\eg, between political parties, or left \vs right) or different stances on politically charged topics.
However, these papers rarely consider how we can make normative judgments on how much different perspectives deserve exposure, the absence of which reinforces the idea that maybe exposure should be equally distributed over them.
Such misguided scholarly position fails to adequately recognize the continuously shifting Overton window of public discourse~\citep{lehman2014brief} and disregards the fact that different political beliefs do not hold equal merit---to treat them otherwise would amount to \textit{algorithmic bothsidism}~\citep{mitra2025search}.

We should not think of this notion of \textit{societal value} (or its corollary, \textit{societal harmfulness}) as an attribute of the commodified information artifact removed from its relational context that can be determined autocratically by the platform and incorporated into our existing notions of relevance or utility of the information---because that would elevate those who wield power over the platform to a position of gatekeeping what constitutes acceptable speech.
Instead, this calls for acknowledging the complex processes of social construction of information and the necessity to reimagine how we can encode spaces for community participation, deliberation, and negotiation within our IR systems that is informed by structural understanding of power relations between dominant and oppressed groups.
The liberal temptation to cast concerns of social justice as a fairness problem also discourages exploration of alternative interventions.
For example, instead of algorithmically fixing the under-representation of women and people of color in image search results for the query ``CEO'', we could reclaim space on the search result page to teach the history of erasure and undervaluing of gendered and racial labor~\citep{mitra2025search}.
The choice here depends on whether our goal is to defend the system against accusations of bias or to materially support the causes of social justice.

Given these concerns, it is unsurprising that there is little evidence that the existing body of fairness research has found meaningful application in dominant search platforms in the context of social justice concerns~\citep{mitra2025search}; and in the absence of meaningful societal impact, the primary outcome of this body of work---which includes significant contributions from industry---seems largely to convince ourselves of our own altruism and ethics-wash the tech industry.
Herein, lies one of the more serious concerns with the liberal frame that in the absence of a clear articulation of our societal goals, any specification of desired system attributes, like fairness, will be ultimately coopted and systemically stripped of its critical interpretation by those who hold social, economic, and political power over our information access platforms to ensure that this research does not in turn threaten their dominance in society.
This is evident in how large corporations editorialize ``Responsible AI'' research coming out of their institutions to ensure that the critical eye of scholarship is always pointed outwards in search for the ``bad actor'' and must never be turned inwards to challenge their concentration of wealth and power or their complicity in the military-industrial complex~\citep{mitra2026why}.
Putting it differently, only the type of research that does not challenge oppressive power---but instead enshrines it---is tolerated, supported, funded, and publicized by the powers that be.
Even if the IR research community produced effective algorithmic interventions, we must reconcile with the reality that we do not hold the power to enforce the deployment of these interventions in real-world systems; in fact we should recognize that our work is happening in the backdrop of state actors and high-net-worth individuals trying to actively capture our information access platforms for their own ideological goals~\citep{mitra2026information}.
The questions of power concentration over our platforms and ecosystems and large power asymmetries within our community spaces are not orthogonal to our research.
They should be intentionally centered in our work to encourage honest critical discourse within the community and development of power-aware research agendas and theories of change.

While I have so far argued my position in the context of fairness research, similar critique is relevant to other concerns of the liberal frame, such as transparency and sustainability.
Transparency in IR has come to denote a body of research that is largely predicated on the goal of convincing users to trust the IR system.
Contrast this with the more critical notion of transparency that \citet{hollanek2023ai} espouses when he writes ``only the sort of transparency that arises from critique---a method of theoretical examination that, by revealing pre-existing power structures, aims to challenge them---can help us produce technological systems that are less deceptive and more just''.
Similarly, concerns of environmental impact in IR have largely been met with proposals for algorithmic and procedural interventions to bring down the cost of training and inference of models---\eg, ~\citet{scells2022reduce}---without adequate consideration of power relationships between different real-world actors, their incentives, and the well-known phenomenon of  Jevons paradox\footnote{\url{https://en.wikipedia.org/wiki/Jevons_paradox}} that are likely to render these well-intentioned strategies ineffectual.
Another relevant example is the emerging literature on model alignment approaches like reinforcement learning from
human feedback (RLHF)~\citep{christiano2017deep, ziegler2019fine} that seems untethered from the concerns of who holds the power to choose the values that models should be aligned with~\citep{mitra2024sociotechnical}.
Lastly, in the EU, the work on ``open search'' has begun~\citep{CERN2025european} in response to the increasing recognition of US Big Tech's oligopolistic control over search platforms.
However, these initiatives---motivated through the nationalist framing of ``digital sovereignty''~\citep{fitzGerald2025digital, djuric2025canada}---point to saliently different notions of openness than what is required to meet the needs of the global south and marginalized communities~\citep{neophytou2026open}.

To be clear, it is not my intention here to claim that all work in fairness, explainability, and related topics strictly adhere to the liberal frame.
I believe that the reality is much more complicated and much of this work has roots in critical scholarship.
But nonetheless it would be accurate to say that this ``first wave'' of IR \& Society research, if we may call it so, has predominantly operationalized liberal framings and engaged inadequately with critical theories and practices.

I also want to be clear that my strong critique of the liberal frame does not imply a wholesale dismissal on my part of the research that has been produced via this frame and its \textit{theoretical} impact.
Quite the contrary, the seeds that were sown at SWIRL 2018~\citep{culpepper2018research} and the FACTS-IR workshop~\citep{olteanu2021facts} brought together a community of societally-motivated researchers and sparked a body of work that serves as the foundation on which we stand today discussing concerns at the intersection of IR \& society.
My own transition from neural IR research to IR \& Society research was against this backdrop and I personally hold immense gratitude towards these scholars and their scholarship that foregrounded societal concerns in IR research.
However, I also steadfastly believe that the liberal frame is saliently incomplete because it is not grounded in critical analysis, structural understanding of systems of oppression, and theories of power.
The time is now ripe to transition to a critical frame if we want to see our work translate to practical societal impact.

\subsection{The critical frame}
\label{sec:frames-critical}
After 2018, the next SWIRL workshop was organized in 2025, where another focus group convened to discuss IR's responsibility to society.
I had the enormous privilege to be invited to SWIRL'25 and the opportunity to participate in this particular focus group.
A central theme of the discussion centered on explicating the values that guide IR research and centering concerns of social justice, emancipation, and democracy in IR~\citep{trippas2025report}.
Related conversations were already ongoing within the IR community, including on the topics of normative design and evaluation of IR systems~\citep{vrijenhoek2023report} and reasserting IR's interdisciplinary roots~\citep{zangerle2025beyond}.
At that time, I had a couple of manuscripts in the preprint stage that also informed some of the conversations; one on explicating sociotechnical imaginaries in IR~\citep{mitra2025search} and another conceptualizing an emancipatory IR framework~\citep{mitra2025emancipatory}, both of which were later published in the Information Retrieval Journal\footnote{\url{https://irrj.org/}} (IRRJ) the same year.
Other recent work that bridges IR with political science and democratic theory---\eg, \citet{vrijenhoek2021recommenders}---were also discussed.
Those crucial conversations was the spark that prompted Dana Mckay, Michael D. Ekstrand, and myself to join with Sanne Vrijenhoek and Maria Murray to organize the workshop titled \textit{Justice, Emancipation, Democracy, and Information Access (JEDI)}\footnote{\url{https://jedi.inertial.science/sigir2026/}}~\citep{mitra2026jedi} at the ACM SIGIR 2026 conference in NAARM / Melbourne, Australia.

In hindsight, the emerging consensus among a subset of the IR community that the field should acknowledge the saliently political nature of our work and re-theorize our goal for ``societal good'' went hand in hand with making the case that IR research needs critical theories and practices.
If SWIRL 2018 can be said to have kicked off the ``first wave'' of IR \& Society research within the liberal frame, then SWIRL 2025 may turn out to be the catalyst for the ``second wave'' that grounds itself in the critical frame.

The critical frame distinguishes itself from the liberal frame in several key aspects, including:
\begin{enumerate}
    \item \textbf{Structural analysis:} It grounds itself in critical theories and structural critiques of systems of oppression.
    \item \textbf{Nondomination:} It adopts a normative position of nondomination and solidarity with the oppressed.
    \item \textbf{Social construction of technology:} It rejects technological determinism and rationality in favor of reimagining and co-constructing anti-oppressive sociotechnical futures.
    \item \textbf{Social construction of knowledge:} It renounces positivist views of information and the traditional role ascribed to IR within capitalist logics of consumption in favor of reframing IR's role within social processes of sense-making and knowledge production.
    \item \textbf{Praxis:} It affirms emancipatory praxis and movement building as research practices.
\end{enumerate}

Next, I discuss each of these aspects in more details.

\subsubsection{Structural analysis}
\label{sec:frames-critical-theory}
Critical theory is a western tradition of thought with roots in the Frankfurt School in the early 20th century~\citep{fuchs2021critical}.
Early critical theorists include Herbert Marcuse, Max Horkheimer, and Theodor W. Adorno.
Critical theory analyzes, with an intent to challenge, the systemic power relations in society.
It posits that knowledge and social structures are saliently shaped by power relations between dominant and oppressed groups, and emphasizes the need to examine intersecting systems of oppression through historical and structural lenses.
Critical information theory~\citep{fuchs2009towards} further combines critical theory with information studies, and situates the study of information in context of relational power asymmetries, domination, and oppression.
\citet{pyati2006critical} argues that Herbert Marcuse's critique of `technological rationality' as a tool of domination is particularly relevant to a critical theoretic analysis of modern information technologies.

Critical IR must be founded on a theoretical basis for situating and studying IR in the context of power relations and systems of oppression.
As the name suggests, critical IR should be grounded in critical theory, but here I use `critical theory' a bit more loosely than alluding to just the Frankfurt School of scholarship.
I am also including other liberatory epistemologies---such as feminist and decolonial theories---as well as other forms of knowledge from activism, political organizing, and movement building.
I am including the many strands of justice-oriented scholarships in information science and in computing---such as from science and technology studies (STS), Fairness, Accountability, and Transparency (FAccT), and human-computer interaction (HCI)---that engage with liberatory epistemologies, including humanistic~\citep{bardzell2016humanistic},  feminist~\citep{hannigan1993feminist, wajcman2004technofeminism, bardzell2010feminist, d2020data}, queer~\citep{light2011hci, kitzie2022advancing}, postcolonial and decolonial~\citep{irani2010postcolonial, ali2016brief, mohamed2020decolonial, jimenez2023holistic}, anti-racist~\citep{ogbonnaya2020critical}, anti-casteist~\citep{vaghela2022interrupting}, anti-ableist~\citep{williams2021articulations}, anti-fascist~\citep{mcquillan2022anti}, post-capitalist~\citep{feltwell2018grand}, and anarchist~\citep{keyes2019human} thought.
The common thread shared between these diverse bodies of scholarship is that they recognize systems of oppression encoded within our social, politicial, economic, cultural, and technological realms, and how they co-constitute broader matrix of domination~\citep{collins2000black}.

The challenge for the IR community is to develop a critical IR research agenda that engages with this rich body of scholarship; and to develop our own theoretical foundation and reflexive muscles equipped with structural analysis to visibilize how IR interacts and intersects with structural oppression. 
For IR to take this leap, we must both reach back to reconnect with our roots in library and information sciences as well as reach beyond our disciplinary boundaries to engage with critical scholarship from both within and outside of computing.

\subsubsection{Nondomination}
\label{sec:frames-critical-nondomination}
A key distinctive feature of critical theory is its normative commitment to nondomination in explicit solidarity with the oppressed.
\citet{horkheimer1972traditional} echoes this ethos by clarifying that the aim of critical theory is not simply to produce knowledge, but to progress the cause of emancipation.
\citet{fuchs2009towards} asserts that critical theory ``\textit{struggles for a classless, nondominative, co-operative, participatory democracy}''.
Consequently, with respect to information studies, \citet{fuchs2009towards} postulate:
\begin{quoting}
    ``\emph{Critical information theory therefore must study not just the role of information and information concepts in society, academia, nature, culture, etc, but how it is related to processes of oppression, exploitation, and domination, which implies a normative judgment in solidarity with the dominated and for the abolishment of domination.}''
\end{quoting}

It is this commitment to nondomination and participatory democracy that I suggest should be the tenet around which we theorize the notion of ``societal good'' within IR.
System behaviors that are desired by the liberal frame---such as fairness and transparency---must then be appraised with respect to whether they help us to sufficiently challenge systems of oppression and move us towards universal emancipation and participatory democracy.
This goal broadens our purview of concern beyond algorithmic mitigations and asks of us to situate our work in broader sociopolitics and power relationships of our world.
It both clarifies the societal change we aspire to affect through our work as well as calls for the development of relevant theories of change for how we envision our work to achieve its desired outcomes.

\subsubsection{Social construction of technology}
\label{sec:frames-critical-scot}
Technological determinism is a reductionist theory that asserts that technology develops following its own internal logic of efficiency and acts as an exogenous force on society and culture causing them to adapt~\citep{heder2021ai}.
We can observe such a worldview  dominant within computing today that presupposes the inevitability of a future shaped by ``AI'' technologies.
This technological deterministic thinking is also reflected by the liberal frame that is reluctant to challenge industry narratives and holds the capitalist logic of market-based decision-making as both legitimate and outside the realm of critique; therefore relegating itself strictly to the role of trying to moderate or delay the harms and inequities caused by these so-claimed ``inevitable'' technologies.

The critical frame rejects technological determinism in favor of recognizing that technology and society mutually shape each other.
It encourages us to challenge technological and instrumental rationality and reclaim our agency to intentionally choose our desired sociotechnical futures.
These visions of desirable futures constitute our sociotechnical imaginaries~\citep{jasanoff2015dreamscapes} and are unsurprisingly shaped by power relations between social groups and capitalist logic of efficiency.
The IR community must reflect on whose sociotechnical imaginaries are granted normative status in our field and intentionally reimagine a liberatory role of information access technologies in realizing nondominative futures.
In doing so, we must also go beyond trying to make oppressive technologies more transparent or fair, and instead redirect our energy to radically redesigning our sociotechnical information systems in service of society's liberatory aspirations.

\subsubsection{Social construction of knowledge}
\label{sec:frames-critical-scok}
The critical frame challenges positivist views on information and the role ascribed to IR within a capitalist model of information consumption.
In IR, it is typical to assume the existence of a corpus of information artifacts (\eg, documents or items) and a set of information needs of users (represented as queries or their current context) that exist independently of the IR system.
The prescribed goal of the IR system is to match the information artifacts (``supply'') with the information needs (``demand'') based on some notion of relevance that the system should estimate automatically.
Relevance here is generally defined in terms of topical match---\ie, does the information artifact satisfy the information need---but may also consider other attributes such as authority and reputation of the information source and factuality of its claims.
IR tasks, in this context, are often represented as pre-curated datasets with a corpus, a set of information needs, and static relevance judgments that guide the development, training, and evaluation of models for relevance estimation.

The social constructionist view of information refutes this perspective and instead sees the corpus, the information needs, the IR system, and society in complex mutually-shaping relations with each other.
Today's IR systems are no longer simple keyword-based lookup tools; they actively model users and information, and shape consumer behavior, political discourse, and culture~\citep{grimmelmann2008google, gillespie2019algorithmically, hallinan2016recommended}.
They are integral to our social processes of sense-making and knowledge production across communities and populations.
The business models of commercial IR platforms and the sociopolitical values of those who control these platforms wield enourmous influence over global information production, not just consumption.
In light of this, the social constructionist perspective challenges the reductionist view of IR as divorced from historical and current sociopolitical context, and instead encourages us to see IR systems as sociotechnical components embedded within dynamic knowledge ecosystems.
This motivates very different views of IR tasks, those that cannot be captured by static datasets in isolation.
Instead, it requires modeling complex information processes in which information is appraised not just based on their relevance or factuality but in the context of power relations within society.
The role of IR then becomes to actively affect certain process outcomes, including growing our collective understanding of our physical and social environments---and our relations with them---and realizing nondominative futures.

\subsubsection{Praxis}
\label{sec:frames-critical-praxis}
Another distinctive feature of the critical IR frame is that it challenges the boundaries of what we traditionally consider as research activities within the field.
Our desired social transformations cannot be realized by theorization and research being conducted in isolation.
Instead, it calls for recognizing praxis (liberatory research + liberatory action) as legitimate and necessary activity within our work.
Examples of such liberatory actions includes:
\begin{enumerate}
    \item \textbf{Building consensus on red lines in our research}, \eg, prohibiting IR research on generative ads and other related technologies whose core objective is to manipulate public opinion at scale.
    \item \textbf{Building counter-structures}, \eg, building public-service IR platforms as alternatives to for-profit proprietary systems, exploring worker-owned cooperatives as alternatives to for-profit corporations, and building alternatives\footnote{\Eg, \url{https://irrj.org/about}} to the for-profit academic publishing ecosystem.
    \item \textbf{Collective organizing and movement building}, \eg, building collectives within the research community, organizing issue-based protests, supporting labor unionization within academic and industry institutions, and participating in broader political movements and collective actions.
    \item \textbf{Sociopolitical consciousness raising within the research community}, \eg, creating spaces where IR community members can engage in critical dialog as a form of liberatory pedagogy~\citep{freire1970pedagogy}, initiating book clubs and reading groups to engage with critical scholarship and build shared understanding, and leveraging different artistic medium like zines to educate and mobilize the community.
    \item \textbf{Public education}, \eg, creating materials and resources that inform and mobilize civil society against oppressive actors who instrumentalize new technologies to entrench their wealth and power and reinforce historical systems of oppression.
    \item \textbf{Diversity, Equity, and Inclusion}, \eg, ensuring that our communities are welcoming and safe spaces for researchers and practitioners from historically underrepresented and marginalized groups and their radical liberatory politics, and actively countering systemic discrimination and marginalization in our community spaces.
    \item \textbf{Community care}, \eg, creating community infrastructure to ensure psychological and financial safety of community members in the face of likely reprisals from those whose power is threatened by our work, and standing in solidarity with our community members who themselves (or their loved ones) are impacted by or are at risk from ongoing human rights violations, immigration violence, displacement, apartheid, war, and genocide.
\end{enumerate}

\section{Concluding thoughts: From \textit{here} to \textit{there}}
\label{sec:conclusion}
Critical IR presents a high-potential direction for theoretical and applied IR research.
The serious risks to our information ecosystems in the emerging era of technofascism~\citep{coeckelbergh2026technofascism} makes investing in this research critically urgent.
The call to center societal, democratic, and emancipatory values in IR research at SWIRL 2025~\citep{trippas2025report} was in recognition of the several ongoing strands of IR research that are already working to incorporate theories from social and political sciences as well as developing our own critical theoretical foundation for the field.
This includes structural critiques of our current approaches to gender-aware information access~\citep{pinney2023much}, critical analysis of the implications of generative AI for IR~\citep{mitra2024sociotechnical}, and questions of how IR can support different forms of democratic governance~\citep{vrijenhoek2021recommenders}.

In my own work~\citep{mitra2025search}, I have argued the need for IR to reject technological determinism and critically examine the sociotechnical imaginaries that shape our research.
I have also advocated for adopting nondomination as a central tenet in our field, offering the following operative definition for emancipatory IR~\citep{mitra2025emancipatory} towards that goal:
\begin{quoting}
    ``\emph{Emancipatory IR is the study and development of information access methods that challenge all forms of human oppression and situates its activities within broader collective emancipatory praxis.}''
\end{quoting}

I theorized about emancipatory IR intentionally to encourage my IR colleagues to center the concerns of universal humanization and liberation from structural forms of oppression---\eg, colonialism, racism, patriarchy, casteism, transphobia, religious persecution, and ableism---in our work; and to discourage the non-performative academic gaze.
But these concerns should not be simply the focus of a subset of the IR community.
At ECIR 2026 IR-for-Good track~\citep{mitra2025ir}, we posed the following rhetorical question:
\begin{quoting}
    ``\textit{Does a special track on IR-for-Good imply that other contributions to the conference are IR-for-Bad?}''.
\end{quoting}

This is an intentional prompt to the IR community to reaffirm that \textit{all} IR research should aspire to realize societal good, and any demands for scientific inquiry that stands contradictory to the wellbeing of society should be vigorously rejected as \citet{belkin1976some} recommended fifty years ago.
To reemphasize, not only should we inject the critical lens in ongoing research on fairness, explainability, and related topics, we should also subject IR research that do not generally make explicit societal arguments---\eg, work on indexing data structures and AI aided information access---to the same critical demands.
Even sub-topics of IR like data structures that may superficially seem detached from societal concerns, may in fact conceal underlying assumptions about centralized platform control and ownership.
My argument here mirrors the perspective of \citet{fuchs2009towards} that our ultimate goal should not be to realize the critical frame as one alternative that coexists with a plurality of other frames, but to realize a unified critical approach to information studies.

To shift the IR community towards critical theories and practices will require pulling together the current strands of ongoing work into a coherent research area.
But more importantly, it will also require raising the sociopolitical consciousness of the community that has historically taken comfort in being portrayed as apolitical.
Any desired transformation to IR technologies must be preceded by transformation of the IR community itself~\citep{mitra2025search}; and we cannot prescribe this transformation top-down---nor from theorists to practitioners---but instead we must build the scaffolding that others can follow for their own liberatory education and development of their conscientização~\citep{freire1970pedagogy}.
We need community spaces that are explicitly politicized and amenable to open and critical conversations about the role of IR in supporting social justice, emancipation, and democracy---as is our goal with the SIGIR 2026 JEDI workshop~\citep{mitra2026jedi}.

Our work in support of this transition will also require developing new community practices.
For example, ECIR 2026 IR-for-Good track~\citep{mitra2025ir} introduced a new requirement for papers to articulate their theories of change and underlying assumptions that must hold true for their desired goals to be realized in practice.
The motivation was that making our theories of change explicit in our scholarship allows for their necessary critique and refinement over time.
Other relevant practices may include requiring papers to provide positionality statements and elaborate on their ethical considerations.
IR would benefit from looking at adjacent communities---such as IS, FAccT, and HCI---for lessons learned in those fields and actively experiment with new practices to continue evolving our research norms.

While throughout this article, I have called on the IR community to center justice and emancipation in our work, I must forewarn that as a community we must reject any false notions of technosolutionism and vigorously challenge the technologists-as-liberator framing~\citep{mitra2026information}.
Instead, we must do the challenging and radical task of reimagining all IR technologies as sociotechnical systems and encode spaces for community participation, collective deliberation, and democratic negotiation within our platforms, and intentionally allow our platforms to be co-opted by the oppressed in service of their liberatory struggles.

My call for incorporating critical theories and practices in IR is also a call for more rigor in our scholarly scientific activities.
Rigorous IR scholarship cannot emerge from a reactionary understanding of our world and the role of information in it.
Rigor in IR, similar to in adjacent disciplines like AI~\citep{olteanu2025rigor}, must be understood as not just methodological rigor but also epistemic, normative, conceptual, reporting, and interpretative rigor.
In summary, the goal of critical IR research is to employ rigorous science in service of realizing Fuchs's ``classless, nondominative, co-operative, participatory democracy''~\citep{fuchs2009towards}.
\section*{Positionality statement}
I am a middle-class Southasian Bengali immigrant and a cisgender man, who grew up in India, got a doctorate in Computer Science in the UK, and is currently working as an independent researcher in Canada.
My lived experiences in India and as an immigrant worker in the USA, the UK, and Canada continue to shape my sociopolitical values and my research.
I worked as an industry researcher in Big Tech for nearly two decades but last year I decided to leave in objection to their harmful impact on society.\footnote{\url{https://disjunctionsmag.com/articles/why-leaving-big-tech/}}
I participate in social justice movement organizing which has deepened my understanding of anti-oppression struggles and traditions of community resistance and solidarity.
I see positionality as an evolving process.
I continue to reflect on how my identity and lived experiences shape my perspectives, values, and research.
\section*{Territorial acknowledgment}
I acknowledge that I authored this article while located in Tiohtià:ke / Mooniyang situated on unceded Indigenous lands and traditional territory of the Kanien’kehá:ka and Anishnaabe Nations.

\section*{Generative AI usage statement}
I hereby declare that no generative AI technologies were used in the writing of this paper.
 \section*{Acknowledgments}
I gratefully acknowledge feedback from Dana Mckay, Michael D. Ekstrand, and Nicola Neophytou on earlier drafts of this article.

\bibliography{references}

@STRING{chi = {Proc. SIGCHI}}

@STRING{sigir = {Proc. SIGIR}}

@STRING{cikm = {Proc. CIKM}}

@STRING{facct = {Proc. FAccT}}

@STRING{chi = {Proc. CHI}}

@STRING{recsys = {Proc. RecSys}}

@String{Computing = "Computing" }

@String{Springer = "Springer-Verlag" }

@inproceedings{diaz2020evaluating,
  title={Evaluating stochastic rankings with expected exposure},
  author={Diaz, Fernando and Mitra, Bhaskar and Ekstrand, Michael D and Biega, Asia J and Carterette, Ben},
  booktitle=cikm,
  pages={275--284},
  year={2020}
}

@inproceedings{belkin1976some,
  title={Some ethical and political implications of theoretical research in information science},
  author={Belkin, NJ and Robertson, SE},
  booktitle={Proceedings of the ASIS Annual Meeting},
  year={1976}
}

@article{ekstrand2022fairness,
  title={Fairness in information access systems},
  author={Ekstrand, Michael D and Das, Anubrata and Burke, Robin and Diaz, Fernando},
  journal={Foundations and Trends{\textregistered} in Information Retrieval},
  volume={16},
  number={1-2},
  pages={1--177},
  year={2022},
  publisher={Emerald Publishing Limited}
}

@inproceedings{asia:equity-of-attention,
	Address = {New York, NY, USA},
	Author = {Biega, Asia J. and Gummadi, Krishna P. and Weikum, Gerhard},
	Booktitle = sigir,
	Pages = {405--414},
	Publisher = {ACM},
	Title = {Equity of Attention: Amortizing Individual Fairness in Rankings},
	Year = {2018}}

@inproceedings{keyes2019human,
  title={Human-computer insurrection: Notes on an anarchist HCI},
  author={Keyes, Os and Hoy, Josephine and Drouhard, Margaret},
  booktitle={Proceedings of the 2019 CHI conference on human factors in computing systems},
  pages={1--13},
  year={2019}
}

@book{jasanoff2015dreamscapes,
  title={Dreamscapes of modernity: Sociotechnical imaginaries and the fabrication of power},
  author={Jasanoff, Sheila and Kim, Sang-Hyun},
  year={2015},
  publisher={University of Chicago Press}
}

@article{anand2022explainable,
  title={Explainable Information Retrieval: A Survey},
  author={Anand, Avishek and Lyu, Lijun and Idahl, Maximilian and Wang, Yumeng and Wallat, Jonas and Zhang, Zijian},
  journal={arXiv preprint arXiv:2211.02405},
  year={2022}
}

@inproceedings{culpepper2018research,
  title={Research frontiers in information retrieval: Report from the third strategic workshop on information retrieval in lorne (swirl 2018)},
  author={Culpepper, J Shane and Diaz, Fernando and Smucker, Mark D},
  booktitle={ACM SIGIR Forum},
  volume={52},
  number={1},
  pages={34--90},
  year={2018},
  organization={ACM New York, NY, USA}
}

@article{zehlike2021fairness,
  title={Fairness in ranking: A survey},
  author={Zehlike, Meike and Yang, Ke and Stoyanovich, Julia},
  journal={arXiv preprint arXiv:2103.14000},
  year={2021}
}

@inproceedings{olteanu2021facts,
  title={FACTS-IR: fairness, accountability, confidentiality, transparency, and safety in information retrieval},
  author={Olteanu, Alexandra and Garcia-Gathright, Jean and de Rijke, Maarten and Ekstrand, Michael D and Roegiest, Adam and Lipani, Aldo and Beutel, Alex and Olteanu, Alexandra and Lucic, Ana and Stoica, Ana-Andreea and others},
  booktitle={ACM SIGIR Forum},
  volume={53},
  number={2},
  pages={20--43},
  year={2021},
  organization={ACM New York, NY, USA}
}

@article{zhang2020explainable,
  title={Explainable recommendation: A survey and new perspectives},
  author={Zhang, Yongfeng and Chen, Xu and others},
  journal={Foundations and Trends{\textregistered} in Information Retrieval},
  volume={14},
  number={1},
  pages={1--101},
  year={2020},
  publisher={Now Publishers, Inc.}
}

@inproceedings{barocas2020not,
  title={When not to design, build, or deploy},
  author={Barocas, Solon and Biega, Asia J and Fish, Benjamin and Niklas, J{\k{e}}drzej and Stark, Luke},
  booktitle={Proceedings of the 2020 Conference on Fairness, Accountability, and Transparency},
  pages={695--695},
  year={2020}
}

@incollection{noble2018algorithms,
  title={Algorithms of oppression},
  author={Noble, Safiya Umoja},
  booktitle={Algorithms of oppression},
  year={2018},
  publisher={New York university press}
}

@inproceedings{pinney2023much,
  title={Much Ado About Gender: Current Practices and Future Recommendations for Appropriate Gender-Aware Information Access},
  author={Pinney, Christine and Raj, Amifa and Hanna, Alex and Ekstrand, Michael D},
  booktitle={Proceedings of the 2023 Conference on Human Information Interaction and Retrieval},
  pages={269--279},
  year={2023}
}

@inproceedings{patro2022fair,
  title={Fair ranking: a critical review, challenges, and future directions},
  author={Patro, Gourab K and Porcaro, Lorenzo and Mitchell, Laura and Zhang, Qiuyue and Zehlike, Meike and Garg, Nikhil},
  booktitle=facct,
  pages={1929--1942},
  year={2022}
}

@article{wang2023survey,
  title={A survey on the fairness of recommender systems},
  author={Wang, Yifan and Ma, Weizhi and Zhang, Min and Liu, Yiqun and Ma, Shaoping},
  journal={ACM Transactions on Information Systems},
  volume={41},
  number={3},
  pages={1--43},
  year={2023},
  publisher={ACM New York, NY}
}

@article{li2023fairness,
  title={Fairness in Recommendation: Foundations, Methods, and Applications},
  author={Li, Yunqi and Chen, Hanxiong and Xu, Shuyuan and Ge, Yingqiang and Tan, Juntao and Liu, Shuchang and Zhang, Yongfeng},
  journal={ACM Transactions on Intelligent Systems and Technology},
  volume={14},
  number={5},
  pages={1--48},
  year={2023},
  publisher={ACM New York, NY}
}

@article{grimmelmann2008google,
  title={The google dilemma},
  author={Grimmelmann, James},
  journal={NYL Sch. L. Rev.},
  volume={53},
  pages={939},
  year={2008},
  publisher={HeinOnline}
}

@incollection{gillespie2019algorithmically,
  title={Algorithmically recognizable: Santorum's Google problem, and Google's Santorum problem},
  author={Gillespie, Tarleton},
  booktitle={The Social Power of Algorithms},
  pages={63--80},
  year={2019},
  publisher={Routledge}
}

@article{hallinan2016recommended,
  title={Recommended for you: The Netflix Prize and the production of algorithmic culture},
  author={Hallinan, Blake and Striphas, Ted},
  journal={New media \& society},
  volume={18},
  number={1},
  pages={117--137},
  year={2016},
  publisher={Sage Publications Sage UK: London, England}
}

@article{lehman2014brief,
  title={A brief explanation of the Overton window},
  author={Lehman, Joseph},
  journal={Mackinac Center for Public Policy},
  year={2014}
}

@article{hollanek2023ai,
  title={AI transparency: a matter of reconciling design with critique},
  author={Hollanek, Tomasz},
  journal={Ai \& Society},
  volume={38},
  number={5},
  pages={2071--2079},
  year={2023},
  publisher={Springer}
}

@article{christiano2017deep,
  title={Deep reinforcement learning from human preferences},
  author={Christiano, Paul F and Leike, Jan and Brown, Tom and Martic, Miljan and Legg, Shane and Amodei, Dario},
  journal={Advances in neural information processing systems},
  volume={30},
  year={2017}
}

@article{ziegler2019fine,
  title={Fine-tuning language models from human preferences},
  author={Ziegler, Daniel M and Stiennon, Nisan and Wu, Jeffrey and Brown, Tom B and Radford, Alec and Amodei, Dario and Christiano, Paul and Irving, Geoffrey},
  journal={arXiv preprint arXiv:1909.08593},
  year={2019}
}

@inproceedings{bardzell2010feminist,
  title={Feminist HCI: taking stock and outlining an agenda for design},
  author={Bardzell, Shaowen},
  booktitle={Proceedings of the SIGCHI conference on human factors in computing systems},
  pages={1301--1310},
  year={2010}
}

@inproceedings{feltwell2018grand,
  title={"Grand visions" for post-capitalist human-computer interaction},
  author={Feltwell, Tom and Lawson, Shaun and Encinas, Enrique and Linehan, Conor and Kirman, Ben and Maxwell, Deborah and Jenkins, Tom and Kuznetsov, Stacey},
  booktitle={Extended Abstracts of the 2018 CHI Conference on Human Factors in Computing Systems},
  pages={1--8},
  year={2018}
}

@article{wajcman2004technofeminism,
  title={Technofeminism. Cambridge: Polity},
  author={Wajcman, Judy},
  year={2004}
}

@article{light2011hci,
  title={HCI as heterodoxy: Technologies of identity and the queering of interaction with computers},
  author={Light, Ann},
  journal={Interacting with computers},
  volume={23},
  number={5},
  pages={430--438},
  year={2011},
  publisher={OUP}
}

@article{bardzell2016humanistic,
  title={Humanistic Hci},
  author={Bardzell, Jeffrey and Bardzell, Shaowen},
  journal={Interactions},
  volume={23},
  number={2},
  pages={20--29},
  year={2016},
  publisher={ACM New York, NY, USA}
}

@inproceedings{irani2010postcolonial,
  title={Postcolonial computing: a lens on design and development},
  author={Irani, Lilly and Vertesi, Janet and Dourish, Paul and Philip, Kavita and Grinter, Rebecca E},
  booktitle={Proceedings of the SIGCHI conference on human factors in computing systems},
  pages={1311--1320},
  year={2010}
}

@incollection{mcquillan2022anti,
  title={Anti-fascist AI},
  author={McQuillan, Dan},
  booktitle={Resisting AI},
  pages={135--148},
  year={2022},
  publisher={Bristol University Press}
}

@article{ali2016brief,
  title={A brief introduction to decolonial computing},
  author={Ali, Syed Mustafa},
  journal={XRDS: Crossroads, The ACM Magazine for Students},
  volume={22},
  number={4},
  pages={16--21},
  year={2016},
  publisher={ACM New York, NY, USA}
}

@inproceedings{vaghela2022interrupting,
  title={Interrupting merit, subverting legibility: Navigating caste in ‘casteless’ worlds of computing},
  author={Vaghela, Palashi and Jackson, Steven J and Sengers, Phoebe},
  booktitle={Proceedings of the 2022 CHI Conference on Human Factors in Computing Systems},
  pages={1--20},
  year={2022}
}

@book{d2020data,
  title={Data feminism},
  author={D'ignazio, Catherine and Klein, Lauren F},
  year={2020},
  publisher={MIT press}
}

@article{mohamed2020decolonial,
  title={Decolonial AI: Decolonial theory as sociotechnical foresight in artificial intelligence},
  author={Mohamed, Shakir and Png, Marie-Therese and Isaac, William},
  journal={Philosophy \& Technology},
  volume={33},
  pages={659--684},
  year={2020},
  publisher={Springer}
}

@incollection{zuboff2023age,
  title={The age of surveillance capitalism},
  author={Zuboff, Shoshana},
  booktitle={Social Theory Re-Wired},
  pages={203--213},
  year={2023},
  publisher={Routledge}
}

@article{williams2021articulations,
  title={Articulations toward a crip HCI},
  author={Williams, Rua M and Ringland, Kathryn and Gibson, Amelia and Mandala, Mahender and Maibaum, Arne and Guerreiro, Tiago},
  journal={Interactions},
  volume={28},
  number={3},
  pages={28--37},
  year={2021},
  publisher={ACM New York, NY, USA}
}

@article{mitra2025search,
  title={Search and Society: Reimagining Information Access for Radical Futures},
  author={Mitra, Bhaskar},
  journal={Information Retrieval Research},
  volume={1},
  number={1},
  pages={47--92},
  year={2025}
}

@incollection{mitra2024sociotechnical,
  title={Sociotechnical Implications of Generative Artificial Intelligence for Information Access},
  author={Mitra, Bhaskar and Cramer, Henriette and Gurevich, Olya},
  booktitle={Information Access in the Era of Generative AI},
  pages={161--200},
  year={2024},
  publisher={Springer}
}

@inproceedings{scells2022reduce,
  title={Reduce, reuse, recycle: Green information retrieval research},
  author={Scells, Harrisen and Zhuang, Shengyao and Zuccon, Guido},
  booktitle={sigir},
  pages={2825--2837},
  year={2022}
}

@article{mitra2025emancipatory,
  title={Emancipatory Information Retrieval},
  author={Mitra, Bhaskar},
  journal={Information Retrieval Research},
  volume={1},
  number={2},
  pages={313--339},
  year={2025}
}

@inproceedings{trippas2025report,
  title={Report from the fourth strategic workshop on information retrieval in lorne (swirl 2025)},
  author={Trippas, Johanne R and Culpepper, J Shane and others},
  booktitle={ACM SIGIR Forum},
  volume={59},
  number={1},
  pages={68},
  year={2025}
}

@book{freire1970pedagogy,
  title={Pedagogy of the Oppressed},
  author={Freire, Paulo},
  publisher={Herder and Herder New York},
  year={1970}
}

@article{djuric2025canada,
  title={Canada’s digital sovereignty push challenges America’s reach},
  author={Djuric, Mickey},
  year={2025},
  journal={Politico Pro}
}

@article{fitzGerald2025digital,
  title={‘Digital sovereignty’: why the EU may be shifting from internet regulation to building homegrown tech},
  author={FitzGerald, Michael},
  year={2025},
  journal={The Conversation}
}

@inproceedings{vrijenhoek2023report,
  title={Report on normalize: The first workshop on the normative design and evaluation of recommender systems},
  author={Vrijenhoek, Sanne and Michiels, Lien and Kruse, Johannes and Starke, Alain and Guerrero, Jordi Viader and Tintarev, Nava},
  booktitle={CEUR Workshop Proceedings},
  volume={3639},
  year={2023},
  organization={CEUR-WS}
}

@inproceedings{vrijenhoek2021recommenders,
  title={Recommenders with a mission: assessing diversity in news recommendations},
  author={Vrijenhoek, Sanne and Kaya, Mesut and Metoui, Nadia and M{\"o}ller, Judith and Odijk, Daan and Helberger, Natali},
  booktitle={chiir},
  pages={173--183},
  year={2021}
}

@inproceedings{ogbonnaya2020critical,
  title={Critical race theory for HCI},
  author={Ogbonnaya-Ogburu, Ihudiya Finda and Smith, Angela DR and To, Alexandra and Toyama, Kentaro},
  booktitle={Proceedings of the 2020 CHI conference on human factors in computing systems},
  pages={1--16},
  year={2020}
}

@article{heder2021ai,
  title={AI and the resurrection of Technological Determinism},
  author={H{\'e}der, Mih{\'a}ly},
  journal={Inform{\'a}ci{\'o}s T{\'a}rsadalom: T{\'a}rsadalomtudom{\'a}nyi Foly{\'o}irat},
  volume={21},
  number={2},
  pages={119--130},
  year={2021}
}

@article{fuchs2009towards,
  title={Towards a critical theory of information},
  author={Fuchs, Christian},
  journal={tripleC: Communication, Capitalism \& Critique. Open Access Journal for a Global Sustainable Information Society},
  volume={7},
  number={2},
  pages={243--292},
  year={2009}
}

@article{neophytou2026open,
  title={Open, to What End? A Capability-Theoretic Perspective on Open Search},
  author={Neophytou, Nicola and Mitra, Bhaskar},
  journal={arXiv preprint arXiv:2603.14584},
  year={2026}
}

@misc{mitra2025ir,
  title={What is IR-for-Good?},
  author={Mitra, Bhaskar and Heuss, Maria},
  year={2025},
  howpublished={\url{https://bhaskar-mitra.github.io/posts/2025/09/01/what-is-ir-for-good/}}
}

@inproceedings{mitra2026jedi,
  title={Justice, Emancipation, Democracy, and Information Access (JEDI): The SIGIR Workshop on Resisting Corporate and Authoritarian Capture of Information Access Platforms},
  author={Mitra, Bhaskar and Mckay, Dana and Ekstrand, Michael D. and Vrijenhoek, Sanne and Murray, Maria},
  booktitle={sigir},
  year={2026}
}

@inproceedings{zobel2018we,
  title={What we talk about when we talk about information retrieval},
  author={Zobel, Justin},
  booktitle={ACM SIGIR Forum},
  volume={51},
  number={3},
  pages={18--26},
  year={2018},
  organization={ACM New York, NY, USA}
}

@book{wu2017attention,
  title={The attention merchants: The epic scramble to get inside our heads},
  author={Wu, Tim},
  year={2017},
  publisher={Vintage}
}

@article{mitra2026information,
  title={Information Access of the Oppressed: Freirean Design for Emancipatory Information Access},
  author={Mitra, Bhaskar and Neophytou, Nicola and Gururaja, Sireesh},
  journal={arXiv preprint arXiv:2601.09600},
  year={2026}
}

@inproceedings{hanna2020towards,
  title={Towards a critical race methodology in algorithmic fairness},
  author={Hanna, Alex and Denton, Remi and Smart, Andrew and Smith-Loud, Jamila},
  booktitle={Proceedings of the 2020 conference on fairness, accountability, and transparency},
  pages={501--512},
  year={2020}
}

@article{pyati2006critical,
  title={Critical theory and information studies: A Marcusean infusion},
  author={Pyati, Ajit K},
  journal={Policy Futures in Education},
  volume={4},
  number={1},
  pages={83--89},
  year={2006},
  publisher={SAGE Publications Sage UK: London, England}
}

@article{mitra2026why,
  title={Why I'm Leaving Big Tech: A Tech Worker's Reflections},
  author={Mitra, Bhaskar},
  journal={Disjunctions},
  number={1},
  year={2026}
}

@inproceedings{zangerle2025beyond,
  title={Beyond Algorithms: Reclaiming the Interdisciplinary Roots of Recommender Systems (BEYOND 2025)},
  author={Zangerle, Eva and Said, Alan and Bauer, Christine},
  booktitle=recsys,
  pages={1360--1361},
  year={2025}
}

@incollection{fuchs2021critical,
  title={What is Critical Theory?},
  author={Fuchs, Christian},
  booktitle={Foundations of critical theory},
  pages={17--51},
  year={2021},
  publisher={Routledge}
}

@article{coeckelbergh2026technofascism,
  title={Technofascism: AI, Big Tech, and the new authoritarianism},
  author={Coeckelbergh, Mark},
  journal={AI \& SOCIETY},
  pages={1--14},
  year={2026},
  publisher={Springer}
}

@book{collins2000black,
  title={Black feminist thought: Knowledge, consciousness, and the politics of empowerment},
  author={Collins, Patricia Hill},
  year={2000},
  publisher={routledge}
}

@article{horkheimer1972traditional,
  title={Traditional and critical theory},
  author={Horkheimer, Max},
  journal={Critical theory: Selected essays},
  volume={188},
  number={243},
  pages={1--11},
  year={1972}
}

@misc{CERN2025european,
    author = {CERN}, 
    title = {European project to make web search more open and ethical}, 
    year = {2025},
    url = {https://home.cern/news/news/computing/european-project-make-web-search-more-open-and-ethical}
}

@article{olteanu2025rigor,
  title={Rigor in AI: Doing Rigorous AI Work Requires a Broader, Responsible AI-Informed Conception of Rigor},
  author={Olteanu, Alexandra and Blodgett, Su Lin and Balayn, Agathe and Wang, Angelina and Diaz, Fernando and Calmon, Flavio du Pin and Mitchell, Margaret and Ekstrand, Michael and Binns, Reuben and Barocas, Solon},
  journal={arXiv preprint arXiv:2506.14652},
  year={2025}
}

@article{hannigan1993feminist,
  title={A feminist paradigm for library and information science},
  author={Hannigan, Jane Anne and Crew, Hilary},
  journal={Wilson Library Bulletin},
  volume={68},
  number={2},
  pages={28--32},
  year={1993},
  publisher={ERIC}
}

@article{jimenez2023holistic,
  title={A holistic decolonial lens for library and information studies},
  author={Jimenez, Andrea and Vannini, Sara and Cox, Andrew},
  journal={Journal of Documentation},
  volume={79},
  number={1},
  pages={224--244},
  year={2023},
  publisher={Emerald Publishing Limited}
}

@article{kitzie2022advancing,
  title={Advancing information practices theoretical discourses centered on marginality, community, and embodiment: Learning from the experiences of lesbian, gay, bisexual, transgender, queer, intersex, and asexual (LGBTQIA+) communities},
  author={Kitzie, Vanessa L and Wagner, Travis L and Lookingbill, Valerie and Vera, Nicolas},
  journal={Journal of the Association for Information Science and Technology},
  volume={73},
  number={4},
  pages={494--510},
  year={2022},
  publisher={Wiley Online Library}
}
\end{document}